\def \SAIT #1 #2 {{\em Mem.\ Soc.\ Astron.\ It.\/} {\bf #1}, #2}
\def \MESS #1 #2 {{\em The Messenger\/} {\bf #1}, #2}
\def \ASTRNACH #1 #2 {{\em Astron. Nach.\/} {\bf #1}, #2}
\def \AAP #1 #2 {{\em Astron. Astrophys.\/} {\bf #1}, #2}
\def \AAL #1 #2 {{\em Astron. Astrophys. Lett.\/} {\bf #1}, L#2}
\def \AAR #1 #2 {{\em Astron. Astrophys. Rev.\/} {\bf #1}, #2}
\def \AAS #1 #2 {{\em Astron. Astrophys. Suppl. Ser.\/} {\bf #1}, #2}
\def \AJ #1 #2 {{\em Astron. J.\/} {\bf #1}, #2}
\def \ANNREV #1 #2 {{\em Ann. Rev. Astron. Astrophys.\/} {\bf #1}, #2}
\def \APJ #1 #2 {{\em Astrophys. J.\/} {\bf #1}, #2}
\def \APJL #1 #2 {{\em Astrophys. J. Lett.\/} {\bf #1}, L#2}
\def \APJS #1 #2 {{\em Astrophys. J. Suppl.\/} {\bf #1}, #2}
\def \APSS #1 #2 {{\em Astrophys. Space Sci.\/} {\bf #1}, #2}
\def \ASR #1 #2 {{\em Adv. Space Res.\/} {\bf #1}, #2}
\def \BAIC #1 #2 {{\em Bull. Astron. Inst. Czechosl.\/} {\bf #1}, #2}
\def \JSQRT #1 #2 {{\em J. Quant. Spectrosc. Radiat. Transfer\/} {\bf #1}, #2}
\def \MN #1 #2 {{\em Mon. Not. R. Astr. Soc.\/} {\bf #1}, #2}
\def \MEM #1 #2 {{\em Mem. R. Astr. Soc.\/} {\bf #1}, #2}
\def \PLR #1 #2 {{\em Phys. Lett. Rev.\/} {\bf #1}, #2}
\def \PASJ #1 #2 {{\em Publ. Astron. Soc. Japan\/} {\bf #1}, #2}
\def \PASP #1 #2 {{\em Publ. Astr. Soc. Pacific\/} {\bf #1}, #2}
\def \NAT #1 #2 {{\em Nature\/} {\bf #1}, #2}
\def \m{\ifmmode M_\odot\else M$_\odot$\fi}
\def \r{\ifmmode R_\odot\else R$_\odot$\fi}

\def \gta {\mathrel{\vcenter
     {\hbox{$>$}\nointerlineskip\hbox{$\sim$}}}}
\def\ni{\noindent}

\def\kms{km~s$^{-1}$}
\def\ergs{erg~s$^{-1}$}

\def\beq{\begin{equation}}
\def\eeq{\end{equation}}
\def\ref{\reference}
\def\gr{$\gamma$-ray}
\def\grb{$\gamma$-ray burst}
\def\grbs{$\gamma$-ray bursts}
\def\ni{$^{56}$Ni}  
\def\co{$^{56}$Co}

\documentstyle{memsait}

\def\PsfigVersion{1.10}
\def\setDriver{\DvipsDriver} 
\ifx\undefined\psfig\else\endinput\fi
\let\LaTeXAtSign=\@
\let\@=\relax
\edef\psfigRestoreAt{\catcode`\@=\number\catcode`@\relax}
\catcode`\@=11\relax
\newwrite\@unused
\def\ps@typeout#1{{\let\protect\string\immediate\write\@unused{#1}}}

\def\DvipsDriver{
	\ps@typeout{psfig/tex \PsfigVersion -dvips}
\def\PsfigSpecials{\DvipsSpecials} 	\def\ps@dir{/}
\def\ps@predir{} }
\def\OzTeXDriver{
	\ps@typeout{psfig/tex \PsfigVersion -oztex}
	\def\PsfigSpecials{\OzTeXSpecials}
	\def\ps@dir{:}
	\def\ps@predir{:}
	\catcode`\^^J=5
}

\def\figurepath{./:}

\def\DoPaths#1{\expandafter\EachPath#1\stoplist}
\def\leer{}
\def\EachPath#1:#2\stoplist{
  \ExistsFile{#1}{\SearchedFile}
  \ifx#2\leer
  \else
    \expandafter\EachPath#2\stoplist
  \fi}
\def\ps@dir{/}
\def\ExistsFile#1#2{%
   \openin1=\ps@predir#1\ps@dir#2
   \ifeof1
       \closein1
   \else
       \closein1
        \ifx\ps@founddir\leer
           \edef\ps@founddir{#1}
        \fi
   \fi}
\def\get@dir#1{%
  \def\ps@founddir{}
  \def\SearchedFile{#1}
  \DoPaths\figurepath
}

\def\@nnil{\@nil}
\def\@empty{}
\def\@psdonoop#1\@@#2#3{}
\def\@psdo#1:=#2\do#3{\edef\@psdotmp{#2}\ifx\@psdotmp\@empty \else
    \expandafter\@psdoloop#2,\@nil,\@nil\@@#1{#3}\fi}
\def\@psdoloop#1,#2,#3\@@#4#5{\def#4{#1}\ifx #4\@nnil \else
       #5\def#4{#2}\ifx #4\@nnil \else#5\@ipsdoloop #3\@@#4{#5}\fi\fi}
\def\@ipsdoloop#1,#2\@@#3#4{\def#3{#1}\ifx #3\@nnil 
       \let\@nextwhile=\@psdonoop \else
      #4\relax\let\@nextwhile=\@ipsdoloop\fi\@nextwhile#2\@@#3{#4}}
\def\@tpsdo#1:=#2\do#3{\xdef\@psdotmp{#2}\ifx\@psdotmp\@empty \else
    \@tpsdoloop#2\@nil\@nil\@@#1{#3}\fi}
\def\@tpsdoloop#1#2\@@#3#4{\def#3{#1}\ifx #3\@nnil 
       \let\@nextwhile=\@psdonoop \else
      #4\relax\let\@nextwhile=\@tpsdoloop\fi\@nextwhile#2\@@#3{#4}}
\ifx\undefined\fbox
\newdimen\fboxrule
\newdimen\fboxsep
\newdimen\ps@tempdima
\newbox\ps@tempboxa
\long\def\fbox#1{\leavevmode\setbox\ps@tempboxa\hbox{#1}\ps@tempdima\fboxrule
    \advance\ps@tempdima \fboxsep \advance\ps@tempdima \dp\ps@tempboxa
   \hbox{\lower \ps@tempdima\hbox
  {\vbox{\hrule height \fboxrule
          \hbox{\vrule width \fboxrule \hskip\fboxsep
          \vbox{\vskip\fboxsep \box\ps@tempboxa\vskip\fboxsep}\hskip 
                 \fboxsep\vrule width \fboxrule}
                 \hrule height \fboxrule}}}}
\fi
\newread\ps@stream
\newif\ifnot@eof       
\newif\if@noisy        
\newif\if@atend        
\newif\if@psfile       
{\catcode`\%=12\global\gdef\epsf@start{
\def\epsf@PS{PS}
\def\epsf@getbb#1{%
\openin\ps@stream=\ps@predir#1
\ifeof\ps@stream\ps@typeout{Error, File #1 not found}\else
   {\not@eoftrue \chardef\other=12
    \def\do##1{\catcode`##1=\other}\dospecials \catcode`\ =10
    \loop
       \if@psfile
	  \read\ps@stream to \epsf@fileline
       \else{
	  \obeyspaces
          \read\ps@stream to \epsf@tmp\global\let\epsf@fileline\epsf@tmp}
       \fi
       \ifeof\ps@stream\not@eoffalse\else
       \if@psfile\else
       \expandafter\epsf@test\epsf@fileline:. \\%
       \fi
          \expandafter\epsf@aux\epsf@fileline:. \\%
       \fi
   \ifnot@eof\repeat
   }\closein\ps@stream\fi}%
\long\def\epsf@test#1#2#3:#4\\{\def\epsf@testit{#1#2}
			\ifx\epsf@testit\epsf@start\else
\ps@typeout{Warning! File does not start with `\epsf@start'.  It may not be a PostScript file.}
			\fi
			\@psfiletrue} 
{\catcode`\%=12\global\let\epsf@percent=
\long\def\epsf@aux#1#2:#3\\{\ifx#1\epsf@percent
   \def\epsf@testit{#2}\ifx\epsf@testit\epsf@bblit
	\@atendfalse
        \epsf@atend #3 . \\%
	\if@atend	
	   \if@verbose{
		\ps@typeout{psfig: found `(atend)'; continuing search}
	   }\fi
        \else
        \epsf@grab #3 . . . \\%
        \not@eoffalse
        \global\no@bbfalse
        \fi
   \fi\fi}%
\def\epsf@grab #1 #2 #3 #4 #5\\{%
   \global\def\epsf@llx{#1}\ifx\epsf@llx\empty
      \epsf@grab #2 #3 #4 #5 .\\\else
   \global\def\epsf@lly{#2}%
   \global\def\epsf@urx{#3}\global\def\epsf@ury{#4}\fi}%
\def\epsf@atendlit{(atend)} 
\def\epsf@atend #1 #2 #3\\{%
   \def\epsf@tmp{#1}\ifx\epsf@tmp\empty
      \epsf@atend #2 #3 .\\\else
   \ifx\epsf@tmp\epsf@atendlit\@atendtrue\fi\fi}

\chardef\psletter = 11 
\chardef\other = 12

\newif \ifdebug 
\newif\ifc@mpute 
\c@mputetrue 

\let\then = \relax
\def\r@dian{pt }
\let\r@dians = \r@dian
\let\dimensionless@nit = \r@dian
\let\dimensionless@nits = \dimensionless@nit
\def\internal@nit{sp }
\let\internal@nits = \internal@nit
\newif\ifstillc@nverging
\def \Mess@ge #1{\ifdebug \then \message {#1} \fi}

{ 
	\catcode `\@ = \psletter
	\gdef \nodimen {\expandafter \n@dimen \the \dimen}
	\gdef \term #1 #2 #3%
	       {\edef \t@ {\the #1}
		\edef \t@@ {\expandafter \n@dimen \the #2\r@dian}%
		\t@rm {\t@} {\t@@} {#3}%
	       }
	\gdef \t@rm #1 #2 #3%
	       {{%
		\count 0 = 0
		\dimen 0 = 1 \dimensionless@nit
		\dimen 2 = #2\relax
		\Mess@ge {Calculating term #1 of \nodimen 2}%
		\loop
		\ifnum	\count 0 < #1
		\then	\advance \count 0 by 1
			\Mess@ge {Iteration \the \count 0 \space}%
			\Multiply \dimen 0 by {\dimen 2}%
			\Mess@ge {After multiplication, term = \nodimen 0}%
			\Divide \dimen 0 by {\count 0}%
			\Mess@ge {After division, term = \nodimen 0}%
		\repeat
		\Mess@ge {Final value for term #1 of 
				\nodimen 2 \space is \nodimen 0}%
		\xdef \Term {#3 = \nodimen 0 \r@dians}%
		\aftergroup \Term
	       }}
	\catcode `\p = \other
	\catcode `\t = \other
	\gdef \n@dimen #1pt{#1} 
}

\def \Divide #1by #2{\divide #1 by #2} 

\def \Multiply #1by #2
       {{
	\count 0 = #1\relax
	\count 2 = #2\relax
	\count 4 = 65536
	\Mess@ge {Before scaling, count 0 = \the \count 0 \space and
			count 2 = \the \count 2}%
	\ifnum	\count 0 > 32767 
	\then	\divide \count 0 by 4
		\divide \count 4 by 4
	\else	\ifnum	\count 0 < -32767
		\then	\divide \count 0 by 4
			\divide \count 4 by 4
		\else
		\fi
	\fi
	\ifnum	\count 2 > 32767 
	\then	\divide \count 2 by 4
		\divide \count 4 by 4
	\else	\ifnum	\count 2 < -32767
		\then	\divide \count 2 by 4
			\divide \count 4 by 4
		\else
		\fi
	\fi
	\multiply \count 0 by \count 2
	\divide \count 0 by \count 4
	\xdef \product {#1 = \the \count 0 \internal@nits}%
	\aftergroup \product
       }}

\def\r@duce{\ifdim\dimen0 > 90\r@dian \then   
		\multiply\dimen0 by -1
		\advance\dimen0 by 180\r@dian
		\r@duce
	    \else \ifdim\dimen0 < -90\r@dian \then  
		\advance\dimen0 by 360\r@dian
		\r@duce
		\fi
	    \fi}

\def\Sine#1%
       {{%
	\dimen 0 = #1 \r@dian
	\r@duce
	\ifdim\dimen0 = -90\r@dian \then
	   \dimen4 = -1\r@dian
	   \c@mputefalse
	\fi
	\ifdim\dimen0 = 90\r@dian \then
	   \dimen4 = 1\r@dian
	   \c@mputefalse
	\fi
	\ifdim\dimen0 = 0\r@dian \then
	   \dimen4 = 0\r@dian
	   \c@mputefalse
	\fi
	\ifc@mpute \then
		\divide\dimen0 by 180
		\dimen0=3.141592654\dimen0
		\dimen 2 = 3.1415926535897963\r@dian 
		\divide\dimen 2 by 2 
		\Mess@ge {Sin: calculating Sin of \nodimen 0}%
		\count 0 = 1 
		\dimen 2 = 1 \r@dian 
		\dimen 4 = 0 \r@dian 
		\loop
			\ifnum	\dimen 2 = 0 
			\then	\stillc@nvergingfalse 
			\else	\stillc@nvergingtrue
			\fi
			\ifstillc@nverging 
			\then	\term {\count 0} {\dimen 0} {\dimen 2}%
				\advance \count 0 by 2
				\count 2 = \count 0
				\divide \count 2 by 2
				\ifodd	\count 2 
				\then	\advance \dimen 4 by \dimen 2
				\else	\advance \dimen 4 by -\dimen 2
				\fi
		\repeat
	\fi		
			\xdef \sine {\nodimen 4}%
       }}

\def\Cosine#1{\ifx\sine\UnDefined\edef\Savesine{\relax}\else
		             \edef\Savesine{\sine}\fi
	{\dimen0=#1\r@dian\advance\dimen0 by 90\r@dian
	 \Sine{\nodimen 0}
	 \xdef\cosine{\sine}
	 \xdef\sine{\Savesine}}}	      

\def\psdraft{
	\def\@psdraft{0}
}
\def\psfull{
	\def\@psdraft{100}
}

\psfull

\newif\if@scalefirst
\def\psscalefirst{\@scalefirsttrue}
\def\psrotatefirst{\@scalefirstfalse}
\psrotatefirst

\newif\if@draftbox
\def\psnodraftbox{
	\@draftboxfalse
}
\def\psdraftbox{
	\@draftboxtrue
}
\@draftboxtrue

\newif\if@prologfile
\newif\if@postlogfile
\def\pssilent{
	\@noisyfalse
}
\def\psnoisy{
	\@noisytrue
}
\psnoisy
\newif\if@bbllx
\newif\if@bblly
\newif\if@bburx
\newif\if@bbury
\newif\if@height
\newif\if@width
\newif\if@rheight
\newif\if@rwidth
\newif\if@angle
\newif\if@clip
\newif\if@verbose
\def\@p@@sclip#1{\@cliptrue}
\newif\if@decmpr
\def\@p@@sfigure#1{\def\@p@sfile{null}\def\@p@sbbfile{null}\@decmprfalse
   \openin1=\ps@predir#1
   \ifeof1
	\closein1
	\get@dir{#1}
	\ifx\ps@founddir\leer
		\openin1=\ps@predir#1.bb
		\ifeof1
			\closein1
			\get@dir{#1.bb}
			\ifx\ps@founddir\leer
				\ps@typeout{Can't find #1 in \figurepath}
			\else
				\@decmprtrue
				\def\@p@sfile{\ps@founddir\ps@dir#1}
				\def\@p@sbbfile{\ps@founddir\ps@dir#1.bb}
			\fi
		\else
			\closein1
			\@decmprtrue
			\def\@p@sfile{#1}
			\def\@p@sbbfile{#1.bb}
		\fi
	\else
		\def\@p@sfile{\ps@founddir\ps@dir#1}
		\def\@p@sbbfile{\ps@founddir\ps@dir#1}
	\fi
   \else
	\closein1
	\def\@p@sfile{#1}
	\def\@p@sbbfile{#1}
   \fi
}
\def\@p@@sfile#1{\@p@@sfigure{#1}}
\def\@p@@sbbllx#1{
		\@bbllxtrue
		\dimen100=#1
		\edef\@p@sbbllx{\number\dimen100}
}
\def\@p@@sbblly#1{
		\@bbllytrue
		\dimen100=#1
		\edef\@p@sbblly{\number\dimen100}
}
\def\@p@@sbburx#1{
		\@bburxtrue
		\dimen100=#1
		\edef\@p@sbburx{\number\dimen100}
}
\def\@p@@sbbury#1{
		\@bburytrue
		\dimen100=#1
		\edef\@p@sbbury{\number\dimen100}
}
\def\@p@@sheight#1{
		\@heighttrue
		\dimen100=#1
   		\edef\@p@sheight{\number\dimen100}
}
\def\@p@@swidth#1{
		\@widthtrue
		\dimen100=#1
		\edef\@p@swidth{\number\dimen100}
}
\def\@p@@srheight#1{
		\@rheighttrue
		\dimen100=#1
		\edef\@p@srheight{\number\dimen100}
}
\def\@p@@srwidth#1{
		\@rwidthtrue
		\dimen100=#1
		\edef\@p@srwidth{\number\dimen100}
}
\def\@p@@sangle#1{
		\@angletrue
		\edef\@p@sangle{#1} 
}
\def\@p@@ssilent#1{ 
		\@verbosefalse
}
\def\@p@@sprolog#1{\@prologfiletrue\def\@prologfileval{#1}}
\def\@p@@spostlog#1{\@postlogfiletrue\def\@postlogfileval{#1}}
\def\@cs@name#1{\csname #1\endcsname}
\def\@setparms#1=#2,{\@cs@name{@p@@s#1}{#2}}
\def\ps@init@parms{
		\@bbllxfalse \@bbllyfalse
		\@bburxfalse \@bburyfalse
		\@heightfalse \@widthfalse
		\@rheightfalse \@rwidthfalse
		\def\@p@sbbllx{}\def\@p@sbblly{}
		\def\@p@sbburx{}\def\@p@sbbury{}
		\def\@p@sheight{}\def\@p@swidth{}
		\def\@p@srheight{}\def\@p@srwidth{}
		\def\@p@sangle{0}
		\def\@p@sfile{} \def\@p@sbbfile{}
		\def\@p@scost{10}
		\def\@sc{}
		\@prologfilefalse
		\@postlogfilefalse
		\@clipfalse
		\if@noisy
			\@verbosetrue
		\else
			\@verbosefalse
		\fi
}
\def\parse@ps@parms#1{
	 	\@psdo\@psfiga:=#1\do
		   {\expandafter\@setparms\@psfiga,}}
\newif\ifno@bb
\def\bb@missing{
	\if@verbose{
		\ps@typeout{psfig: searching \@p@sbbfile \space  for bounding box}
	}\fi
	\no@bbtrue
	\epsf@getbb{\@p@sbbfile}
        \ifno@bb \else \bb@cull\epsf@llx\epsf@lly\epsf@urx\epsf@ury\fi
}	
\def\bb@cull#1#2#3#4{
	\dimen100=#1 bp\edef\@p@sbbllx{\number\dimen100}
	\dimen100=#2 bp\edef\@p@sbblly{\number\dimen100}
	\dimen100=#3 bp\edef\@p@sbburx{\number\dimen100}
	\dimen100=#4 bp\edef\@p@sbbury{\number\dimen100}
	\no@bbfalse
}
\newdimen\p@intvaluex
\newdimen\p@intvaluey
\def\rotate@#1#2{{\dimen0=#1 sp\dimen1=#2 sp
		  \global\p@intvaluex=\cosine\dimen0
		  \dimen3=\sine\dimen1
		  \global\advance\p@intvaluex by -\dimen3
		  \global\p@intvaluey=\sine\dimen0
		  \dimen3=\cosine\dimen1
		  \global\advance\p@intvaluey by \dimen3
		  }}
\def\compute@bb{
		\no@bbfalse
		\if@bbllx \else \no@bbtrue \fi
		\if@bblly \else \no@bbtrue \fi
		\if@bburx \else \no@bbtrue \fi
		\if@bbury \else \no@bbtrue \fi
		\ifno@bb \bb@missing \fi
		\ifno@bb \ps@typeout{FATAL ERROR: no bb supplied or found}
			\no-bb-error
		\fi
		\count203=\@p@sbburx
		\count204=\@p@sbbury
		\advance\count203 by -\@p@sbbllx
		\advance\count204 by -\@p@sbblly
		\edef\ps@bbw{\number\count203}
		\edef\ps@bbh{\number\count204}
		\if@angle 
			\Sine{\@p@sangle}\Cosine{\@p@sangle}
	        	{\dimen100=\maxdimen\xdef\r@p@sbbllx{\number\dimen100}
					    \xdef\r@p@sbblly{\number\dimen100}
			                    \xdef\r@p@sbburx{-\number\dimen100}
					    \xdef\r@p@sbbury{-\number\dimen100}}
                        \def\minmaxtest{
			   \ifnum\number\p@intvaluex<\r@p@sbbllx
			      \xdef\r@p@sbbllx{\number\p@intvaluex}\fi
			   \ifnum\number\p@intvaluex>\r@p@sbburx
			      \xdef\r@p@sbburx{\number\p@intvaluex}\fi
			   \ifnum\number\p@intvaluey<\r@p@sbblly
			      \xdef\r@p@sbblly{\number\p@intvaluey}\fi
			   \ifnum\number\p@intvaluey>\r@p@sbbury
			      \xdef\r@p@sbbury{\number\p@intvaluey}\fi
			   }
			\rotate@{\@p@sbbllx}{\@p@sbblly}
			\minmaxtest
			\rotate@{\@p@sbbllx}{\@p@sbbury}
			\minmaxtest
			\rotate@{\@p@sbburx}{\@p@sbblly}
			\minmaxtest
			\rotate@{\@p@sbburx}{\@p@sbbury}
			\minmaxtest
			\edef\@p@sbbllx{\r@p@sbbllx}\edef\@p@sbblly{\r@p@sbblly}
			\edef\@p@sbburx{\r@p@sbburx}\edef\@p@sbbury{\r@p@sbbury}
		\fi
		\count203=\@p@sbburx
		\count204=\@p@sbbury
		\advance\count203 by -\@p@sbbllx
		\advance\count204 by -\@p@sbblly
		\edef\@bbw{\number\count203}
		\edef\@bbh{\number\count204}
}
\def\in@hundreds#1#2#3{\count240=#2 \count241=#3
		     \count100=\count240	
		     \divide\count100 by \count241
		     \count101=\count100
		     \multiply\count101 by \count241
		     \advance\count240 by -\count101
		     \multiply\count240 by 10
		     \count101=\count240	
		     \divide\count101 by \count241
		     \count102=\count101
		     \multiply\count102 by \count241
		     \advance\count240 by -\count102
		     \multiply\count240 by 10
		     \count102=\count240	
		     \divide\count102 by \count241
		     \count200=#1\count205=0
		     \count201=\count200
			\multiply\count201 by \count100
		 	\advance\count205 by \count201
		     \count201=\count200
			\divide\count201 by 10
			\multiply\count201 by \count101
			\advance\count205 by \count201
		     \count201=\count200
			\divide\count201 by 100
			\multiply\count201 by \count102
			\advance\count205 by \count201
		     \edef\@result{\number\count205}
}
\def\compute@wfromh{
		\in@hundreds{\@p@sheight}{\@bbw}{\@bbh}
		\edef\@p@swidth{\@result}
}
\def\compute@hfromw{
	        \in@hundreds{\@p@swidth}{\@bbh}{\@bbw}
		\edef\@p@sheight{\@result}
}
\def\compute@handw{
		\if@height 
			\if@width
			\else
				\compute@wfromh
			\fi
		\else 
			\if@width
				\compute@hfromw
			\else
				\edef\@p@sheight{\@bbh}
				\edef\@p@swidth{\@bbw}
			\fi
		\fi
}
\def\compute@resv{
		\if@rheight \else \edef\@p@srheight{\@p@sheight} \fi
		\if@rwidth \else \edef\@p@srwidth{\@p@swidth} \fi
}
\def\compute@sizes{
	\compute@bb
	\if@scalefirst\if@angle
	\if@width
	   \in@hundreds{\@p@swidth}{\@bbw}{\ps@bbw}
	   \edef\@p@swidth{\@result}
	\fi
	\if@height
	   \in@hundreds{\@p@sheight}{\@bbh}{\ps@bbh}
	   \edef\@p@sheight{\@result}
	\fi
	\fi\fi
	\compute@handw
	\compute@resv}
\def\OzTeXSpecials{
	\special{empty.ps /@isp {true} def}
	\special{empty.ps \@p@swidth \space \@p@sheight \space
			\@p@sbbllx \space \@p@sbblly \space
			\@p@sbburx \space \@p@sbbury \space
			startTexFig \space }
	\if@clip{
		\if@verbose{
			\ps@typeout{(clip)}
		}\fi
		\special{empty.ps doclip \space }
	}\fi
	\if@angle{
		\if@verbose{
			\ps@typeout{(rotate)}
		}\fi
		\special {empty.ps \@p@sangle \space rotate \space} 
	}\fi
	\if@prologfile
	    \special{\@prologfileval \space } \fi
	\if@decmpr{
		\if@verbose{
			\ps@typeout{psfig: Compression not available
			in OzTeX version \space }
		}\fi
	}\else{
		\if@verbose{
			\ps@typeout{psfig: including \@p@sfile \space }
		}\fi
		\special{epsf=\ps@predir\@p@sfile \space }
	}\fi
	\if@postlogfile
	    \special{\@postlogfileval \space } \fi
	\special{empty.ps /@isp {false} def}
}
\def\DvipsSpecials{
	\special{ps::[begin] 	\@p@swidth \space \@p@sheight \space
			\@p@sbbllx \space \@p@sbblly \space
			\@p@sbburx \space \@p@sbbury \space
			startTexFig \space }
	\if@clip{
		\if@verbose{
			\ps@typeout{(clip)}
		}\fi
		\special{ps:: doclip \space }
	}\fi
	\if@angle
		\if@verbose{
			\ps@typeout{(clip)}
		}\fi
		\special {ps:: \@p@sangle \space rotate \space} 
	\fi
	\if@prologfile
	    \special{ps: plotfile \@prologfileval \space } \fi
	\if@decmpr{
		\if@verbose{
			\ps@typeout{psfig: including \@p@sfile.Z \space }
		}\fi
		\special{ps: plotfile "`zcat \@p@sfile.Z" \space }
	}\else{
		\if@verbose{
			\ps@typeout{psfig: including \@p@sfile \space }
		}\fi
		\special{ps: plotfile \@p@sfile \space }
	}\fi
	\if@postlogfile
	    \special{ps: plotfile \@postlogfileval \space } \fi
	\special{ps::[end] endTexFig \space }
}
\def\psfig#1{\vbox {
	\ps@init@parms
	\parse@ps@parms{#1}
	\compute@sizes
	\ifnum\@p@scost<\@psdraft{
		\PsfigSpecials 
		\vbox to \@p@srheight sp{
			\hbox to \@p@srwidth sp{
				\hss
			}
		\vss
		}
	}\else{
		\if@draftbox{		
			\hbox{\fbox{\vbox to \@p@srheight sp{
			\vss
			\hbox to \@p@srwidth sp{ \hss 
			 \hss }
			\vss
			}}}
		}\else{
			\vbox to \@p@srheight sp{
			\vss
			\hbox to \@p@srwidth sp{\hss}
			\vss
			}
		}\fi

	}\fi
}}
\psfigRestoreAt
\setDriver
\let\@=\LaTeXAtSign

\begin{opening}
\title{THE SUPERNOVA GAMMA-RAY BURST CONNECTION} 
\author{J. CRAIG WHEELER$^1$, PETER H\"OFLICH$^1$, LIFAN WANG$^1$}
\institute{$^1$Department of Astronomy, University of Texas, Austin, Texas}
\date{} 
\end{opening}

\begin{document}

\oddpagefooter{}{}{} 
\evenpagefooter{}{}{} 
\ 
\bigskip

\begin{abstract}
Study of the polarization of supernovae has suggested that the
core collapse process may be intrinsically strongly asymmetric.
There is a tentative trend for supernova with smaller 
envelopes showing more polarization, with SN~Ic having the smallest
envelopes and showing the largest polarization.
The recent discovery of the unusual supernova SN~1998bw and its apparent 
correlation with the gamma-ray burst GRB~980425 has raised new issues 
concerning both the \grbs\ and supernovae. 
SN~1998bw resembled a SN~Ic, but was unusually bright at maximum 
light in the optical and radio, and its expansion velocities were large.
This makes SN~1998bw a possible candidate for a ``hypernova" 
with explosion energies exceeding $10^{52} erg$.  
We show that the light curve  of SN1998bw can be understood as the result 
viewing an aspherical explosion roughly along the symmetry 
axis of an exploding, non-degenerate C/O core of a massive star with
a kinetic energy of $2 \times 10^{51}$ erg,  a total ejecta 
mass of $2 M_\odot$, and a $^{56}Ni$ mass of $0.2M_\odot$. 
In this model, the high expansion velocities are a direct 
consequence of the aspherical explosion which, in turn, 
produces oblate iso-density  contours and that accounts for the polarization. 
It is not yet clear how either the hypernovae or these asymmetric
models can produce \grbs.
\end{abstract} 

\section{Introduction}
 
Due to its correlation in time and location, 
the $\gamma$-ray burst GRB~980425
has a high probability of being associated with SN~1998bw (Galama et al. 1998).
This connection is supported by the association of
a radio source with SN~1998bw  that requires rapid expansion
(Waxman \& Loeb 1998) and probably relativistic expansion
(Kulkarni et al. 1998).  Recent corrections to
the BeppoSAX positions and record of time variability
show that the supernova falls in the error box (Piro, et al. 1998)
of a time-variable X-ray source (Pian, et al. 1998).  
From optical spectra, SN~1998bw was classified as a 
SN~Ic by Patat and Piemonte (1998).  
What sets SN~1998bw apart from other Type Ic (SN~Ic) are higher expansion 
velocities as indicated by the Si II and Ca H and K lines
($\approx$ 30 to 50$\%$ higher at maximum light than  SN1994I and SN1983V;
Clochiatti  \& Wheeler 1997), the red colors at maximum light, and the 
large intrinsic brightness.  A peak luminosity of 
$1.3\pm 0.6\times10^{43}$ \ergs can be inferred from the redshift 
of the host galaxy ($z=0.0085$, Tinney et al. 1998) and the reddening
($A_V=0.2^m$, Schlegel et al. 1998), if we assume $H_o = $ 67 km/s/Mpc.
The uncertainties are rather large as the host galaxy is not
yet fully in the Hubble flow, so  the peculiar velocity may 
be of the order of 300 to 400 \kms,  $H_o$ is known
only to an accuracy of $\approx 10 \%$, and 
 $A_V$  may vary by $\approx 0.1^m$.
The radio source associated with SN~1998bw is the brightest ever
observed for a supernova (Kulkarni, et al. 1998).
 
The properties of SN~1998bw  suggest that it was a ``hypernova" event 
(Paczy\'nski 1997). Based on their light curve calculations,
Iwamoto et al. (1998) and Woosley, Eastman \& Schmidt (1998) d
erived explosion energies of
20-50 foe and 22 foe ($1 foe = 10^{51} erg$)
ejecta masses of 12-15 $M_\odot$ and 6 $M_\odot$, 
and $^{56}$Ni masses of 0.6-0.8 and 0.5 $M_\odot$, respectively.

Guided by the deduced properties of more traditional core 
collapse supernovae and SN~Ic in particular, 
we have investigated whether asphericity can provide an 
alternative explanation for SN~1998bw that allows its optical
properties, at least, to remain in the range of ``normal" SN~Ic.

\section{Polarization of Supernovae}

Both spectral analyses and 
light curve calculations support the picture that
SN~II, SN~Ib and SN~Ic form a sequence involving core 
collapse with successively smaller H and He envelopes
(Clochatti \& Wheeler 1997).
The analysis of spectra and light curves gives, however, essentially no
insight into the geometry of the expanding envelope. 
Polarization, on the other hand, provides a unique tool to explore asymmetries.
For the last several years we have engaged in a program to obtain
spectropolarimetry of as many supernovae as possible from
McDonald Observatory (Wang et al. 1996).
Linear polarization of $\approx 1 \% $ seems to be typical for SN~II
(Wang, Wheeler \& H\"oflich 1998).  
There is a trend, however, for the observed polarization 
to increase in core-collapse supernovae with decreasing envelope mass,
e.g. from SN~II to SN~Ic (Wang et al. 1998).  
SN 1987A, with a hydrogen envelope of about 10\m, 
had a polarization of $\approx 0.5 \%$; for SN~1993J, 
with a hydrogen envelope mass of only a few tenths of \m,
the observed linear polarization was as high as $\approx 1.5 \% $; 
for the SN~Ic 1997X, with no hydrogen and little or no helium 
envelope, the polarization was even higher, perhaps several percent 
(Wang et al. 1998; Wang \& Wheeler 1998). 
The suggestion is that the closer one looks to the collapsing core
itself, the larger the polarization and the larger the asymmetry.
This trend, while tentative, clearly points toward the 
interpretation that the explosion itself is strongly asymmetric. 
 
From theoretical calculations of scattering dominated atmospheres, 
this size of polarization, $\gta 1 $ \%,
requires axis ratios of the order of 2 or 3 to 1, 
requiring the ejecta to be highly aspherical. 
The  luminosity $L(\Theta )$ will vary by a factor of
 $\approx $ 2 as the line of sight varies from the equator to the pole 
(H\"oflich 1991, H\"oflich et al. 1995).  
 
Given the ubiquitous presence of polarization in core collapse
supernovae and especially SN~Ic,  inclusion of asphericity 
effects in SN~Ib/c may prove to be critical to their
understanding (Wang et al. 1998). 
Polarization was been observed in SN1998bw (Kay et al. 1998)
at the level of about $0.5\%$.  Asymmetries thus cannot be
neglected in a complete model for this event.  An obvious question
is whether the asymmetries are directly connected to 
the fact that it produced a \grb.

\section{Type Ib/c Supernovae and Gamma-Ray Bursts}

Wang \& Wheeler (1998) examined the question of whether
there is a general connection between supernovae and \grbs. 
They concluded that SN~Ia could be excluded from any such
correlation at the $4\sigma$ level, that the sample of
sufficiently well-studied SN~II was too small to reach
meaningful conclusions, but that a correlation with
SN~Ib/c could not be ruled out.  Table I gives the list of
candidate SN~Ib/c -- \grb\ correlations suggested by
Wang \& Wheeler.

                  
\vspace{1cm} 
\centerline{\bf Table 1 - Supernovae of Type Ib/c and GRB Associations}
\begin{table}[h]
\hspace{1.5cm} 
\begin{tabular}{|c|c|c|c|c|}
\hline
SN & Type & Dates & RA (2000.0) & DEC (2000.0)  \\ 
\hline

SN 1994at &  Ib/c   & $\sim$961009 & 17.1   & -1.0 \\
GRB 960925 &        &  & 29.37   & -13.9 \\
\hline
SN 1996N & Ib      & $\sim$960310 & 54.7   & -26.3 \\
GRB 960221&        &               & 47.8   & -31.24 \\
\hline
SN 1997B& Ic       & $\sim$970104 & 88.3   & -17.9 \\
GRB 971218&        &    & 97.75   & -21.73 \\
\hline
SN 1997dq& Ib       & $\sim$971105 & 175.2 & +11.5 \\
GRB 971013&        &    & 167.0   & +2.7 \\
\hline
SN 1997ef& ?       & $\sim$971205 & 119.3   & +49.6 \\
GRB 971152&        &   & 84.6   & +41.7 \\
\hline
SN 1997ei& Ic      & $\sim$971223 & 178.5  & +58.5 \\
GRB 971120&        &   & 155.7  & +76.4 \\
\hline
SN 1998bw &  ?     & 980424-980427 & 19 35 03.3   & -52 50 44.8\\
GRB 980425&        & 980425.909    & 19 35 21     & -52 52 19\\

\hline
\end{tabular}
\end{table}

Kippen et al. (1998) found no evidence for a connection between
supernovae and \grbs\ and Graziani, Lamb, \& Marion (1998) concluded
that not all SN~Ib/c could be associated with \grbs, although a
fraction might be.  With the small sample of SN~Ib/c and
partial sky coverage, it will be difficult to resolve this issue
with statistics alone. Woosely, Eastman, \& Schmidt (1998) 
suggested that SN~1998cy might be associated with GRB~9970514.  
Unlike the
suggestions in Table I, SN~1998cy shows both narrow and broad
features of hydrogen.  This event was even brighter than SN~1998bw
(Schmidt, 1998).  If it is correlated with a \grb\ it represents
another type of exploding object.  When it was first observed,
SN~1997ef was noted to be an object of unprecedented spectral
features, although it bore some resemblance to SN~Ic.  The
similarity of SN~1997ef to SN~1998bw was pointed out by
Garnavich (1997).  Wang \& Wheeler found that SN~1997ef was
just beyond the $2\sigma$ BATSE error box for GRB~971115.
A model for the light curve of SN~1997ef has been presented
by Iwamoto et al. (1999) based on a carbon/oxygen core exploding with
a ``normal" kinetic energy.  At this meeting, Nomoto reported
a revised calculation based on a ``hypernova" scenario with
in excess of 10 foe of kinetic energy that better matched
the line profiles.  The question of whether this event was
polarized and hence asymmetric also arises.  Garnavich, Jha, \&
Kirshner (1998) report that SN 1998ey was ``identical" to
SN~19978ef two weeks after discovery and similar to SN~1998bw.
They report no correlation with a BATSE trigger, but again,
BATSE only surveys a portion of the sky at one instant and so
the constraint on a single event with the uncertainty in
explosion much greater than the orbit of CGRO is weak.
  
\section{Description of the Concept and  Numerical Methods}

We have investigated the question of whether the optical
properties of SN~1998bw can be explained based on an
asymmetric explosion with otherwise normal amounts of
kinetic energy and total ejecta and \ni\ mass as opposed to
the unprecedentedly large values of these quantities
required in the ``hypernova" models of Iwamoto et al. (1998)
and Woosely, Eastman, \& Schmidt (1998).
 
For the initial setup, we use the chemical and density 
structures of spherical C/O cores of Nomoto and Hashimoto (1988). 
These structures are scaled to adjust the total mass of the ejecta.
This is an approximation, but the details 
of the chemical profiles are not expected to effect the light curves. 
 
The explosion models are calculated using a one-dimensional
radiation-hydro code which includes  a detailed nuclear network.
The code also simultaneously solves for the radiation 
transport via moment equations.  Photon redistribution and
thermalization is based on detailed NLTE-models. Several hundred frequency 
groups are used to calculate  monochromatic light curves, 
frequency-averaged Eddington factors, and opacity means.
A Monte Carlo scheme is used for $\gamma$-rays.
For details, see H\"oflich, Wheeler \& Thielemann 
(1999) and references therein.

Aspherical density structures are constructed based on the 
original spherical density distribution.  For the simple models presented
here,  we impose the asymmetry after the ejecta has
reached the homologous expansion phase.  We generate an
asymmetric configuration by preserving the mass fraction
per steradian from the spherical model, but imposing a different
law of homologous expansion as a function of the angle $\Theta$
from the equatorial plane.
For typical density structures, a higher energy deposition 
along the polar axis results in oblate density structures.
Such an energy pattern may be produced if jet-like structures 
are formed during the central core collapse as suggested by 
Wang \& Wheeler (1998). In contrast, a prolate density structure
would be produced if more energy is released in the equatorial region
than in the polar direction. For more details, 
see H\"oflich, Wang \& Wheeler (1999).
 
The emissivity along a given isodensity contour in the
asymmetric models is taken to
be constant and equal  to the corresponding equivalent velocity-weighted
mean layer in the spherical model.
The bolometric and broad band light curves are constructed 
by convolving the spherical light curves with the photon redistribution 
functions, L$(\Theta)$/L(mean), that are calculated
by the Monte Carlo code for polarization 
(H\"oflich 1991, H\"oflich et al. 1995).
Typical conditions at the photosphere, and therefore the colors,
are expected to be similar in both the spherical and aspherical models
because the energy flux, $F(\Theta)$, is found to be similar both 
in the spherical and aspherical configuration to within $\approx 40 \% $. 
Stationarity is assumed to calculate the photon redistribution functions
since the geometry does not change during the typical diffusion 
time scale.  We assume implicitly the same
mean diffusion time scales for both the spherical and aspherical 
configurations.  This mostly effects the very early phases of the 
light curve when the hydrodynamical time scales are short. For more details, 
see H\"oflich, Wang \& Wheeler (1998).

\section{Results}
 
We construct models in such a way a way that at day 20
the axis ratio at the photosphere is 2.
In comparison to the spherical model, the homology expansion parameters are 
a factor of $\approx 2.2 $ larger along the pole for oblate ellipsoids 
and a factor of $\approx 1.5$ larger for prolate ellipsoids 
along the equator.
 
We first calculated aspherical light curves based 
on the C/O core CO21  which gives a good representation of 
the BVRI light curves of the SN~Ic 1994I (Iwamoto et al. 1994). 
The ejecta mass is $0.8 M_\odot $ and the explosion energy is
$E_{kin} = 10^{51} erg $. A mass of $0.08~ M_\odot $ of $^{56}Ni$ is ejected.
This model failed  to produce the peak brightness by a factor of 2, 
gave too short a rise time by about 5 days, and shows
blue color indices at maximum light.

To boost the total luminosity to the level of observation 
we increased the amount of ejected  $^{56}Ni$ to 0.2 $M_\odot$. 
This quantity of nickel is still below the estimates for 
$^{56}Ni$ of 0.3 $M_\odot$ in the bright SN~II 1992am (Schmidt et al. 1997). 
As shown in Fig.1,  the time of maximum light is 
rather insensitive to asphericity effects.
The need to delay the time to maximum and to produce  
the red color at maximum light suggested the need to increase 
the ejecta mass with an appropriate increase in the 
kinetic energy to provide the observed expansion.
We thus computed a series of models with $M_{ej}=2 M_\odot$ 
$E_{kin}=2\times10^{51}$ erg and $M_{Ni} = 0.2$\m. 
 \begin{figure}[t]
\psfig{figure=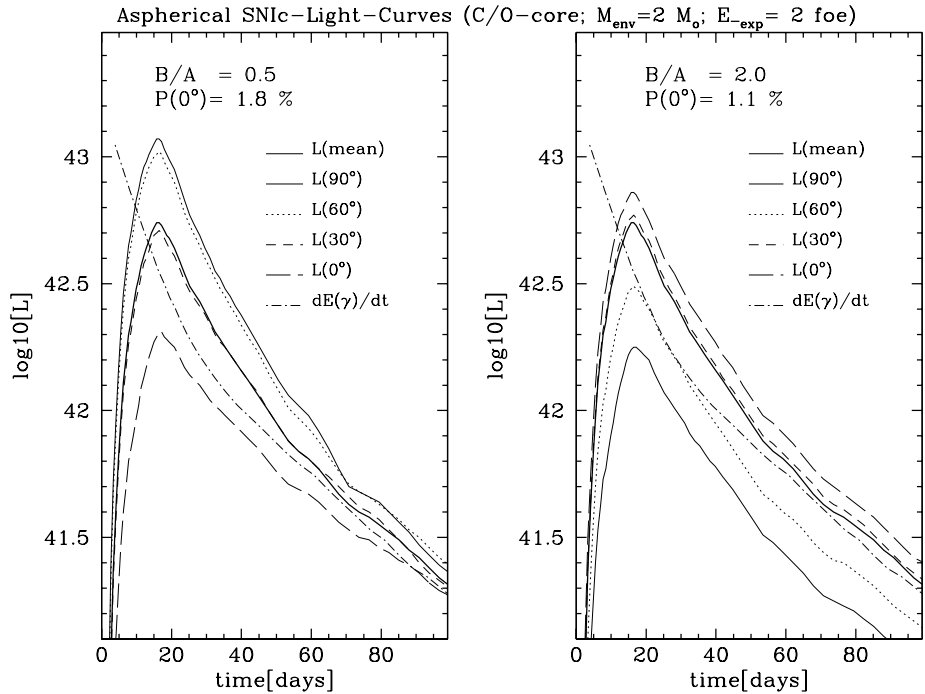,width=13.0cm,rwidth=12.0cm,clip=,angle=270}
\caption{Directional dependence of the bolometric light curve 
for oblate (left) and prolate (right) ellipsoids. 
The luminosity of the corresponding spherical model is shown as $L(mean)$.
In addition, the instantaneous $\gamma-ray$ deposition is shown.
$P(0^o)$ is the polarization at maximum light 
($ P(\Theta) \approx P(0^o) \times cos^2\Theta$).}
 \end{figure}
 
Asphericity of the amplitude assumed here
can change the luminosity over a range of roughly 2 magnitudes (Fig. 1).
For oblate ellipsoids, the luminosity is
enhanced along the pole whereas for prolate density structures 
the enhancement occurs in the equatorial direction. 
Combined with the polarization properties, 
this provides a clear separation between oblate and prolate geometries as 
$P$ always goes to 0 if the structure is seen pole-on and $P$  
increases towards lower latitudes (H\"oflich 1991). 

 \begin{figure}[t]
\psfig{figure=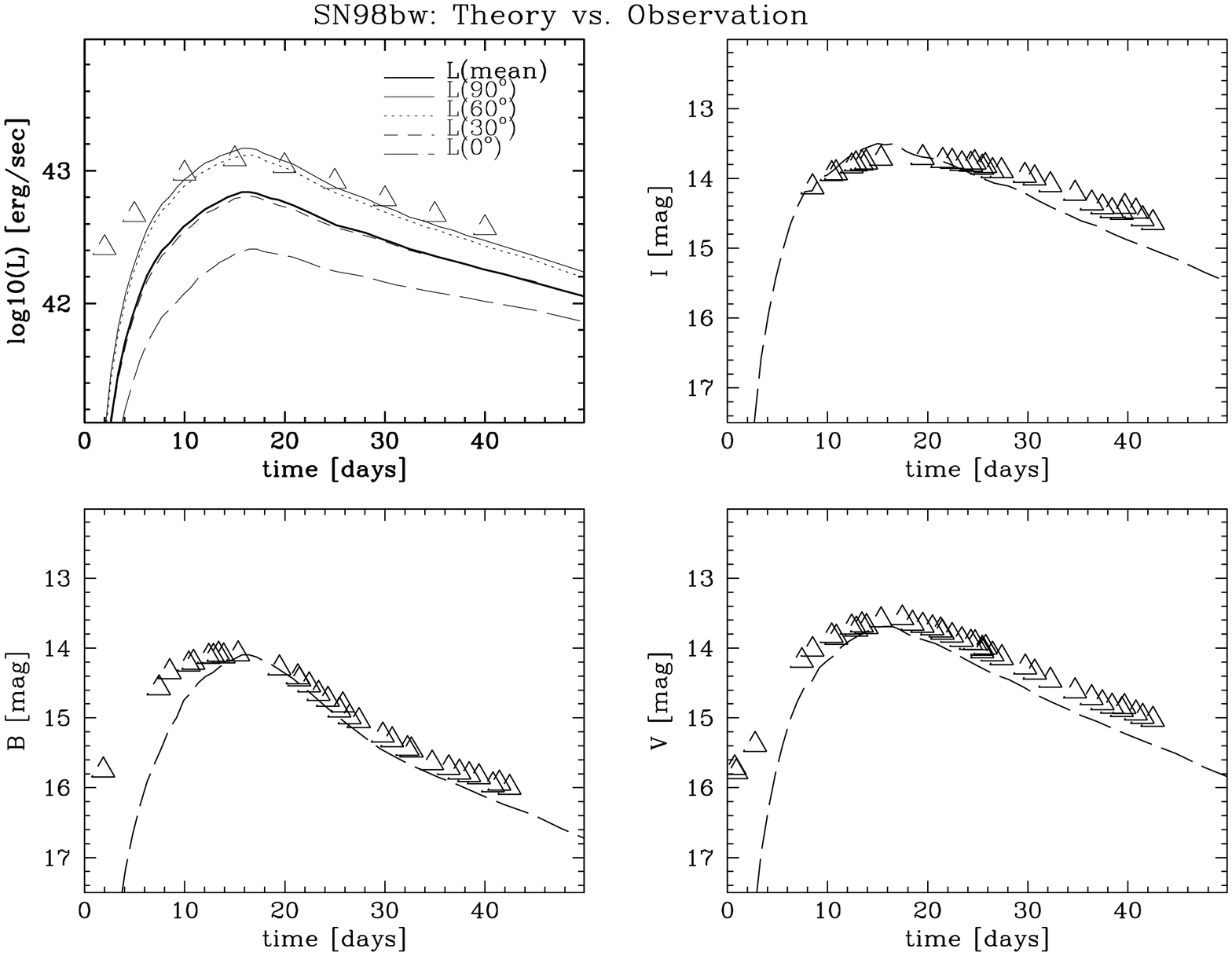,width=13.0cm,rwidth=12.0cm,clip=,angle=270}
\caption{Comparison of bolometric and broad-band  light curves 
as observed for SN1998bw with those of an oblate ellipsoid seen 
at high angle with respect to the equator assuming a distance of 
36 Mpc and  $A_V=0.2^m$.}
 \end{figure}

Observations of the polarization of SN~1998bw 23 days after
the explosion show little polarization ($< 1 \% $, Patat et al. 1998). 
By day 58, the intrinsic polarization was reported to be $0.5 \%$ 
(Kay et al. 1998).  Polarization data on SN~Ic is rare, 
but this value is less than seen in some SN~Ic and related events
(see above). SN~1998bw was also rather bright.
This combination implies oblate geometries if asymmetry is involved.

For the comparison between the observed and theoretical light curves
as shown in Fig. 2,
we have used the relative calibration of Woosley, Eastman,
\& Schmidt (1998)
for the ``bolometric light curve".  The broad-band data was obtained from 
Galama et al (1998).  To account for the obervations with
our asymmtric model, the object
must be seen from an angle of $\ge 60^o$ from the equator. 
The same conclusion can be drawn independently from 
the  detected, but relatively small linear polarization.
This model thus suggests that the maximum polarization
of SN~1998bw was greater than the $0.5 \%$ measured by Kay et al. (1998). 
Overall, the broad-band light curves agree with the data 
within the uncertainties.
The intrinsic color excess B-V matches the observations within $0.1^m$ 
and, after the initial rise of $\approx 7$ days, the agreement in each band
is better than $0.3^m$. 
The main discrepancy with the observations occurs during 
the initial rise when the diffusion time scales are
much longer than the expansion time.  Under these conditions, 
our approximation for redistribution of the energy
of a spherical model breaks down since the diffusion time scale 
is long compared to the hydrodynamical time scale.
The decline after maximum is slightly too steep in the models both in the 
bolometric and broad band light curves.
This is likely to be related to the energy generation in the 
envelope by \gr\ deposition or to the change in the escape 
probability of low energy photons. The decline rate immediately after peak 
can be reduced by increasing the amount of radioactive $^{56}Ni$ by
$\approx 40 \%$. In our models, the escape probability for $\gamma$-rays 
increased rapidly between day 20 and 80 from 3\% to 50\%.
An alternative means to flatten the light curve is to reduce the increase in 
escape probability. This can be achieved by a 
modification restricted to the inner layers of the ejecta 
because the escape probability is determined by those layers.
Either the expansion velocity of the inner layers can be 
reduced or the density gradient might be steeper. 
Both are expected for strongly aspherical explosions.

\section{Discussion and Conclusions}
 
The ``hypernovae" models for SN~1998bw provide interesting fits
to the data, but these models have some problems.  
Although the fit of Iwamoto et al. (1998) 
of the light curve is excellent with 
errors $\leq 0.3^m $ over 40 days, the spectra
show absorption lines that are too narrow by a factor of 
2 to 3 indicating  too narrow a range of formation in velocity space. 
This may be related to the high envelope mass.
In the lower mass models of Woosley, Eastman, \& Schmidt (1998),
the computed color indices (B-V, V-R, V-I) are too red at all epochs.
The hypernova models presented to date are spherically
symmetric and hence make no pretense of accounting for
the observed polarization, but this is a critical feature
and cannot be ignored.

We have shown that the high apparent luminosity of SN~1998bw 
may be understood within the framework of ``classical" SNIc
by invoking asymmetry of the ejecta of the degree required to
account for the polarization.  

Even with the invocation of significant asymmetry, our model for 
SN~1998bw remains at the bright end of the scale for normal
SN~Ic.  We note that the luminosity of 
SN~1998bw may be uncertain by a factor of 2 due to non-Hubble
motion within the cluster and uncertainties in the 
Hubble constant and reddening. For a model with an ejected mass
of 2 $M_\odot$, an explosion energy of $2\times 10^{51}$ erg, and 
a $^{56}Ni$- ejection of $0.2 M_\odot$,
both the bolometric and broad-band light curves 
are rather well reproduced by an oblate 
ellipsoid with an axis ratio of 2 to 1 which is observed within 
$30\deg$ of the symmetry axis.
This angle for the line of sight is consistent with the low 
(but still significant) polarization observed for SN~1998bw. 
In a Lagrangian frame, the polar expansion velocity is a factor of 2 larger
than the mean velocity. This is also in agreement with the rather 
large expansion velocities seen in SN~1998bw.
 
Neither this model nor the hypernova models give an obvious explanation of
the \grb\ nor the especially bright radio emission. 
Woosley, Eastman, \& Schmidt (1998) have analyzed the possibility of 
$\gamma $-ray bursts in the framework
of spherical models. Even with their explosion energies of more 
than 20 foe they showed that the \grb\ associated with
SN~1998bw/GRB~980425 cannot be explained by the acceleration
of matter to relativistic speeds at shock-breakout. 
In our picture, the specific energy released in the polar region is comparable.
We want to stress, however, that the asymmetry in the energy distribution 
in the model presented here is set after homologeous expansion 
has been established. Because the early hydrodynamical evolution will tend 
to wipe out asymmetries, the initial anisotropy in the energy distribution
as the explosion is initiated after core collapse
is expected to be significantly higher. 
 
We have shown that SN~1998bw may be 
understood within the framework of ``classical" core-collapse supernovae 
rather than by a ``hypernova," but the actual model 
parameters must be regarded as uncertain both because of the 
model assumptions and the uncertainty in the observed luminosity.
In light of the good fits, however, SN~1998bw may indeed be a ''hypernova."
Continuous measurements of the polarization and the velocity 
of $^{56}Co$ lines are critical to unreveal the nature and 
geometry of this object.  If the late-time tail tracks the 
\co\ decay line, then the ejected \ni\ mass could be determined.
This might prove the simplest discriminant between the
``hypernova" models and models based on significant asymmetry.

There remains the broader issue of
the possible connection of supernovae and \grbs.
If some of the \grbs\ at large redshifts are driven by supernovae,
then the associated energy flow must be strongly collimated and aspherical
core collapse may be a natural way to produce such jets.
The basic purpose of this work is to underline the fact that asymmetries must
be taken into account in a complete consideration of core
collapse supernovae, of SN~1998bw in particular, and, by extension
of \grbs.
 
\acknowledgements

We thank  Ken Nomoto for providing us with the monochromatic light curve 
data in digital form and Titus Galama, Brian Schmidt, Ken Nomoto,
Stan Woosely, Martin Rees, Ralph Wijers, Dieter Hartmann, and
Peter M\'esz\'aros for valuable conversations. 
This research was supported in part by 
NSF Grant AST 9528110, NASA Grant NAG 5-2888, and a grant 
from the Texas Advanced Research Program.







\end{document}